\documentclass{ws-mpla}

\begin{document}

\markboth{Andr\'e de Gouv\^ea}
{Neutrinos Have Mass -- So What?}

%
\catchline{}{}{}{}{}
%

\title{\bf NEUTRINOS HAVE MASS -- SO WHAT?}

\author{\footnotesize ANDR\'E DE GOUV\^EA}

\address{Physics \& Astronomy Department, Northwestern University, 2145 Sheridan Road\\
Evanston, IL 60208-3112,
USA\\
degouvea@northwestern.edu}

\maketitle

\pub{Received (Day Month Year)}{Revised (Day Month Year)}

\begin{abstract}
In this brief review, I discuss the new physics unveiled by neutrino oscillation experiments over the past several years, and discuss several attempts at understanding the mechanism behind neutrino masses and lepton mixing. It is fair to say that, while significant theoretical progress has been made, we are yet to construct a coherent picture that naturally explains non-zero, yet tiny, neutrino masses and the newly revealed, puzzling patterns of lepton mixing. I discuss what the challenges are, and point to the fact that more experimental input (from both neutrino and ``non-neutrino'' experiments) is dearly required -- and that new data is expected to reveal, in the next several years, new information. Finally, I draw attention to the fact that neutrinos may have only just begun to reshape fundamental physics, given the fact that we are still to explain the LSND anomaly and because the neutrino oscillation phenomenon is ultimately  sensitive to very small new-physics effects.

\end{abstract}

\ccode{PACS Nos.: 14.60.Pq,11.30.Fs,11.30.Hv,12.15.Ff}

\section{Surprise in the Neutrino Sector}	
\label{intro}

Over the past six years, an astonishing sequence of experimental results have revealed,\cite{solar,atm,accelerator,reactor,lsnd} without a shadow of a doubt, that neutrinos change flavor. This means that a neutrino produced in a well-defined weak eigenstate $\nu_{\alpha}$ can be detected in a distinct weak eigenstate $\nu_{\beta}$\footnote{Weak eigenstates are labeled by the flavor of the charged lepton involved in the charged current weak process responsible for either producing or detecting neutrinos, {\it i.e.}, $\alpha,\beta=e,\mu,\tau,\ldots$.} after propagating a macroscopic distance. Furthermore, it has been established that the rate of flavor change $P_{\alpha\beta}$ depends on the neutrino energy $E_{\nu}$ and on the propagation distance $L$.

The simplest (and only completely satisfactory) way to explain the observed neutrino flavor changing phenomena is to postulate that neutrinos have distinct, nonzero masses, and that the neutrino mass eigenstates\footnote{Mass eigenstates are eigenstates of the free particle Hamiltonian in vacuum, {\it i.e.}, $H|\nu_i\rangle=m_i|\nu_i\rangle$ in the neutrino rest frame. These are referred to as $\nu_i$, $i=1,2,3,\ldots$.} are different from the neutrino weak eigenstates. This being the case, neutrinos will undergo neutrino oscillations as they propagate, and $P_{\alpha\beta}$ is the oscillation probability. Besides $L$ and $E_{\nu}$, $P_{\alpha\beta}$ is a function of the neutrino mass-squared differences $\Delta m^2_{ij}\equiv m_j^2-m_i^2$ and the elements of the lepton mixing matrix $U$. In the weak basis where the charged current and the charged lepton masses are diagonal, the lepton mixing matrix is the unitary matrix that relates the neutrino weak eigenstates to the neutrino mass eigenstates: $\nu_{\alpha}=U_{\alpha i}\nu_i$.  

All neutrino data,\cite{solar,atm,accelerator,reactor} with the exception of the LSND anomaly,\cite{lsnd} are explained by three flavor neutrino oscillations. I will comment on the LSND anomaly in section \ref{conclusions}, and, for simplicity and the sake of the presentation, will ignore its existence henceforth. 

For three neutrino flavors, the elements of the lepton mixing matrix are defined by
\begin{equation}
\left(\begin{array}{c}\nu_e \\ \nu_{\mu} \\ \nu_{\tau} \end{array} \right) =\left(\begin{array}{ccc} U_{e1} & U_{e2} & U_{e3} \\ U_{\mu1} & U_{\mu2} & U_{\mu3} \\ U_{\tau1} & U_{e\tau2} & U_{\tau3}\end{array}\right)
\left(\begin{array}{c}\nu_1 \\ \nu_2 \\ \nu_3 \end{array} \right),
\label{UMNS}
\end{equation}
and are, of course, not all independent. It is customary\cite{pdg} to parameterize $U$ in Eq.~(\ref{UMNS}) with three mixing angles $\theta_{12},\theta_{13},\theta_{23}$ and three complex phases, $\delta,\xi,\zeta$, defined by
\begin{equation}
\frac{|U_{e2}|^2}{|U_{e1}|^2}\equiv \tan^2\theta_{12};
~~~~\frac{|U_{\mu3}|^2}{|U_{\tau3}|^2}\equiv \tan^2\theta_{23};~~~~
U_{e3}\equiv\sin\theta_{13}e^{-i\delta},
\end{equation}
with the exception of $\xi$ and $\zeta$, the so-called Majorana CP-odd phases. These are only physical if the neutrinos are Majorana fermions, and have, unfortunately, virtually no effect in flavor-changing phenomena. We have no idea what their values are or whether they are physical observables, and I will ignore them henceforth, unless otherwise noted.

In order to relate the mixing elements to experimental observables, it is necessary to define the neutrino mass eigenstates, {\it i.e.}, to ``order'' the neutrino masses. This will be done in the following way: $m_2^2>m_1^2$ and $\Delta m^2_{12}<|\Delta m^2_{13}|$. In this case, there are three mass-related observables: $\Delta m^2_{12}$ (positive definite), $|\Delta m^2_{13}|$, and the sign of $\Delta m^2_{13}$. A positive sign for $\Delta m^2_{13}$ implies $m_3^2>m_2^2$ -- a so-called normal mass-hierarchy -- while a negative sign for $\Delta m^2_{13}$ implies $m_3^2<m_1^2$  -- a so-called inverted mass-hierarchy. 

Detailed combined analyses of all neutrino data are consistent, at the three sigma confidence level, with\cite{valle_new}
\begin{itemize}
\item $\sin^2\theta_{12}=0.30\pm0.08$, mostly from solar and KamLAND data;
\item $\sin^2\theta_{23}=0.50\pm0.18$, mostly from atmospheric neutrino data;
\item $\sin^2\theta_{13}\leq 0047$, mostly from atmospheric and Chooz\cite{chooz} data;
\item $\Delta m^2_{12}=(8.1\pm1.0)\times 10^{-5}$~eV$^2$, mostly from solar and KamLAND data;
\item $|\Delta m^2_{13}|=(2.2\pm1.1)\times 10^{-3}$~eV$^2$, mostly from atmospheric neutrino data.
\end{itemize}
I refer readers to Ref.~\refcite{valle_new} for details of the analyses and the results, including correlations, etc. We still need to find out 
\begin{itemize}
\item the sign of $\Delta m^2_{13}$ or what is the neutrino mass hierarchy?;
\item the value of $\theta_{13}$ or is $|U_{e3}|\neq 0$?;
\item the value of $\delta$ or is there CP-invariance violation in neutrino oscillations?;
\item the sign of $1/2-\sin^2\theta_{23}$ or is atmospheric mixing really maximal?
\end{itemize}

The goal of this brief review is {\sl not} to explain how the above measurements have been performed, nor to discuss the several strategies for determining the currently unknown oscillation parameters. I will, instead, concentrate on discussing the importance of the results summarized above for fundamental high energy physics and our understanding of Nature at very small distance scales. 

The most important piece of information we obtained thanks to the herculean effort of the several experiments cited above can be summarized in a single sentence: NEUTRINOS HAVE MASS. And it turns out that massive neutrinos constitute the only palpable evidence we have that the standard model of electroweak and strong interactions (SM) does not describe all strong, electromagnetic and weak phenomena.

Along with this, two other rather puzzling features have been identified as potential clues of what lies beyond. One is that, while neutrino masses turn out to be nonzero, they are exceedingly tiny. 

The most straight forward (and less model dependent) limit on the absolute value of the neutrino masses is provided by precise studies of the beta-decay spectrum of tritium. These reveal,\cite{pdg} at the 95\% confidence level, that $m_{\beta}^2\equiv\sum_i|U_{ei}|^2m^2_i<5.0$~eV$^2$, which implies, given that the neutrino mass-squared differences are much smaller than 1~eV$^2$, that $\forall i$, $|m_i|<2.2$~eV at the 95\% confidence level. More stringent, but also more model dependent, bounds can be obtained from cosmological data,\cite{cosmic_nus} and for searches for neutrino-less double beta decay.\cite{pdg,0nubb} By combining all bounds, it is safe to say that all neutrinos weigh less than 1~eV. 

\begin{figure}[th]
\centerline{\psfig{file=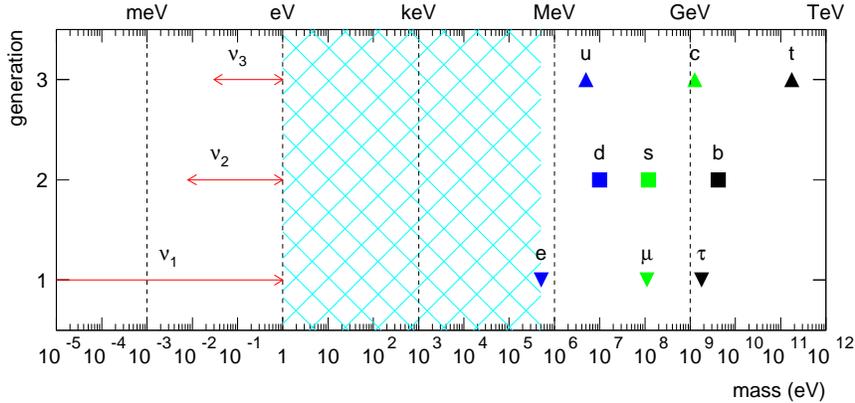,width=0.9\textwidth}}
\vspace*{8pt}
\caption{Masses of all known fundamental fermions.}
\label{allmasses}
\end{figure}

Figure \ref{allmasses} depicts the value of the masses of all known fundamental fermions. Note the log scale. Fermion masses are very hierarchical -- it takes over thirteen orders of magnitude to fit them all in one plot! Quark masses span  five orders of magnitude, while charged fermion masses span over three orders of magnitude. We don't know why fermion masses are distributed in this way. 

Another remarkable fact (the one I am most interested in here) is the fact that the ratio of the largest possible neutrino mass to the lightest known charged fermion mass (the electron mass) is at least one order of magnitude less than the ratio of the electron mass to the top quark mass. Furthermore, while the electron--top ``gap'' is populated by all other charged leptons and quarks, the heaviest-neutrino--electron ``gap'' seems to be deserted.  We also don't know why this is the case, but it does seem to be Nature's way of saying that there is something special about neutrino masses. It may be that neutrino masses are qualitatively different from charged fermion masses.

 The other peculiar fact revealed by the neutrino oscillation data is that lepton mixing turns out to be quite distinct from quark mixing. I'll return to this issue in section \ref{section:mixing}.

Before proceeding, I'll briefly describe the SM, and what is meant, here, by going beyond it. The SM is a Lorentz invariant quantum field theory, and its renormalizable Lagrangian is uniquely determined once one specifies its internal symmetries (gauged $SU(3)_c\times SU(2)_L\times U(1)_Y$ invariance) and particle content ($Q,u,d,L,e$, the matter fields, plus $H$, the Higgs doublet scalar field). The fact that the Lagrangian is renormalizable implies that, in principle, the SM is valid up to arbitrarily high energy scales.\footnote{This is not the case. Gravitational interactions are not accounted for in the SM. This is usually a very good approximation, which breaks down, naively, if one tries to describe processes that occur around the Planck scale $M_{\rm Pl}\sim 10^{18}$~GeV. I'll argue that neutrino masses may be interpreted as evidence that the SM is not an appropriate description for physical processes at significantly lower energy scales, and will ignore ``quantum gravity'' concerns.} It is easy to check that given the SM as I defined it above, neutrinos are strictly massless. I'll argue in the next two sections that neutrino masses can be included by modifying the SM appropriately, and that the modification changes {\sl qualitatively} what is meant by the SM.  

\section{Majorana Neutrinos}

There are several ways of modifying the SM and allowing nonzero neutrino masses. The amount of experimental information available is, however, still too small to allow a particular candidate ``new SM'' to be chosen over another, but there is reason to believe that more information is on the way in the near (few years) to intermediate (several years) future. Here, I'll very briefly describe some of these candidates.

The, arguably, simplest way to ``add'' neutrino masses to the SM is to give up on the renormalizability of the Lagrangian. This allows one to add (an infinite number of) irrelevant operators consistent with the symmetries:
\begin{equation}
{\cal L}_{\rm new}={\cal L}_{\rm SM}-\lambda^{\alpha\beta}\frac{L_{\alpha}HL_{\beta}H}{2M}+{\cal O}\left(\frac{1}{M^2}\right).
\label{eq:majorana}
\end{equation}
$L$ is the left-handed lepton double field. Note that, unless otherwise noted, all fermion fields are understood to be Weyl fermions, such that, for example, the charged lepton Yukawa operator is written as $LHe$, where $e$ is the (positively) charged lepton singlet field. 

Two facts are remarkable. One is that $(LH)^2$ is the {\sl only} dimension-five operator allowed by the SM gauge invariance and particle content.\cite{weinberg} The other is that, as long as $M$ is much larger than $\langle H\rangle$, the Higgs 
vacuum expectation value, the only observable consequence of Eq.~(\ref{eq:majorana})\footnote{This is not necessarily correct. One also has to worry about dimension six operators that lead to baryon number violation and a finite lifetime for the proton.} is that neutrinos get a nonzero  mass after electroweak symmetry breaking: $m_{\nu}=\lambda\langle H\rangle^2/M$. An extra ``bonus'' is that neutrino masses are naturally much smaller than all other fermion masses $m_f\propto\langle H\rangle$ by a factor $\langle H\rangle/M$.

Another important consequence of Eq.~(\ref{eq:majorana}) is that lepton number is not a good symmetry ($(LH)^2$ breaks lepton number by two units). Lepton number violation is ``encoded'' in the fact that the neutrinos are Majorana fermions. 

It remains to discuss the ``new physics'' scale $M$. It can be described, roughly, as the energy scale above which  
Eq.~(\ref{eq:majorana}) is no longer valid. There is very little information regarding the magnitude of $M$, but one can set an upper bound for $M$ by assuming that $\lambda\sim 4\pi$,\cite{willen} {\it i.e.}, by assuming that the physics replacing Eq.~(\ref{eq:majorana}) at the scale $M$ is strongly coupled. In this case 
\begin{equation}
M\lesssim4\pi\frac{\langle H\rangle^2}{m_{\nu}}\sim10\frac{100^2~\rm GeV^2}{10^{-10}~\rm GeV}\left(\frac{100~\rm meV}{m_{\nu}}\right)=10^{15}~\rm GeV\left(\frac{100~\rm meV}{m_{\nu}}\right).
\end{equation}
If Eq.~(\ref{eq:majorana}) is indeed the correct low-energy description of Nature, neutrino masses represent the first direct evidence that the SM is an effective field theory, valid up to an energy scale less that $10^{15}$~GeV (or so), which is, in turn, much less than the Planck scale.\cite{willen,degouvea_valle} It is also impressive that the upper bound above coincides, qualitatively, with the energy scale where all three running gauge coupling constants of the SM seem to meet, {\it i.e.}, the grand unified scale $M_{\rm GUT}\sim 10^{15-16}$~GeV.

There are several different proposals for the physics that replaces Eq.~(\ref{eq:majorana}). The arguably simplest and most elegant one is the seesaw mechanism,\cite{seesaw} described by the following Lagrangian
\begin{equation}
{\cal L}_{\rm seesaw}={\cal L}_{\rm SM}-y_{\alpha i}L_{\alpha}HN_i-\frac{M_N^{ij}}{2}N_iN_j+H.c.~.
\label{eq:seesaw}
\end{equation} 
$N_i$ are SM singlet fermion fields, and Eq.~(\ref{eq:seesaw}) is the most general renormalizable Lagrangian consistent with gauge invariance and the addition of these new fields to the SM particle content. Note that the presence of $N_i$ breaks lepton number since these have Majorana masses $M_N$. In the limit $M_N\gg\langle H\rangle$, one can integrate out these ``right-handed neutrino'' fields in order to describe phenomena at and below the weak scale. The effective Lagrangian turns out to be 
\begin{equation}
{\cal L}_{\rm new}={\cal L}_{\rm SM}-\left(y^{\rm t}M_N^{-1}y\right)^{\alpha\beta}L_{\alpha}HL_{\beta}H+{\cal O}\left(\frac{1}{M_N^2}\right).
\label{eq:majorana2}
\end{equation}
Eq.~(\ref{eq:majorana2}) is the same as Eq.~(\ref{eq:majorana}) after one equates $y^tM_N^{-1}y=\lambda/2M$. Eq.~(\ref{eq:seesaw}) is, therefore, an ultraviolet completion of Eq.~(\ref{eq:majorana}), where $M$ is associated with the mass of the right-handed neutrino fields. 

Majorana neutrino masses are the only observable physical consequence of Eq.~(\ref{eq:seesaw}), with the possible important exception of the decay of right-handed neutrino fields in the very early universe. These may have lead to the observed baryon asymmetry of the universe, generated via the lepton asymmetry which naturally arises from CP-violating and lepton-number violating decays of the $N_i$.\cite{leptogenesis}  I refer readers to Ref.~\refcite{leptogenesis_review} for recent, detailed studies of this so-called leptogenesis mechanism. Finally, it is worthwhile to mention that Eq.~(\ref{eq:seesaw}), combined with weak-scale supersymmetry, may lead to observable rates for charged lepton flavor violating processes.\cite{clfv_seesaw}

The new physics scale $M$ need not be as high as the GUT scale. For example, a rather different way of introducing Majorana neutrino masses is to assume weak-scale supersymmetry (SUSY), including the violation of invariance under R-parity. For concreteness, let us add to the minimal supersymmetric standard model (MSSM) superpotential the trilinear R-parity violating term $\lambda^{\alpha\beta i}\bf{L}_{\alpha}\bf{L}_{\beta}\bf{E}_i$ ({\bf L}, {\bf E} are superfields). After SUSY and electroweak symmetry breaking, neutrinos get a Majorana mass at the one-loop level\cite{rpv_numass}
\begin{equation} 
m_{\nu}\sim\frac{\lambda^2}{16\pi^2}\left(\frac{m_{\ell}m_{\ell^{\prime}}}{m_{\tilde{\ell}^{\prime\prime}}}\right),
\end{equation}
where $\ell$ ($\tilde{\ell}$) are charged leptons (sleptons). Assuming $m_{\tilde{\ell}}$ are around 1~TeV, neutrino masses are due to new physics at the electroweak breaking scale. Experimental constraints require the magnitudes of the relevant $\lambda$ couplings to be quite small -- making this procedure not particularly elegant -- but, on the other hand, the phenomenology of these types of models is rather rich. This is especially true of expectations for rare charged lepton processes, including $\mu^+\to e^+\gamma$, $\mu^+\to e^+e^-e^+$, and $\mu\to e$-conversion in nuclei.\cite{clfv_rpv} In particular, expectations for the ratios of  the expected rates for $\mu^+\to e^+e^-e^+$ and $\mu\to e$-conversion in nuclei to $\mu^+\to e^+\gamma$ are very different from more canonical ``seesaw-induced'' charged-lepton flavor violation.

\section{Dirac Neutrinos}

A completely different option is to assume that neutrinos are, similar to all charged matter fields, Dirac fermions. In this case, in order to render the neutrinos massive, it suffices to add extra SM gauge singlet Weyl fermions $N_i$ (``right-handed neutrinos''), and Yukawa couplings between $H$, $L_{\alpha}$, $N_i$:
\begin{equation}
{\cal L}_{\rm new}={\cal L}_{\rm SM}-y_{\alpha i}L_{\alpha}HN_i+H.c.~.
\label{eq:dirac}
\end{equation} 
After electroweak symmetry breaking, the neutrino mass matrix is given by $m_{\nu}=y\langle H\rangle$, similar to the up-type and down-type quark mass matrices and the charged lepton mass matrix. The magnitude of the neutrino masses requires $y\lesssim 10^{-12}$, at least six orders of magnitude smaller than the electron Yukawa coupling. It is clear that a natural explanation for the smallness of the neutrino mass is {\sl not} contained in Eq.~(\ref{eq:dirac}). On the positive side, Eq.~(\ref{eq:dirac}) is renormalizable, meaning that this new SM version is, naively, valid up to arbitrarily high energy scales. 

Modulo a natural explanation for the size of the neutrino mass, one could argue that Eq.~(\ref{eq:dirac}) is a rather innocuous addition to the SM Lagrangian. This is not correct, and he reason for it is somewhat subtle. Eq.~(\ref{eq:dirac}) is not the most general, renormalizable Lagrangian consistent with the symmetries of the SM. Once the fields $N_i$ are introduced, the dimension-three Majorana mass operators $\frac{1}{2}M_N^{ij}N_iN_j$ should also have been introduced, given the fact that $N_i$'s are gauge singlets, {\it cf.} Eq.~(\ref{eq:seesaw}). One needs, therefore, to modify the {\sl symmetry structure} of the SM in order to forbid a Majorana mass term for the right-handed neutrinos. One simple way of doing this is to add to the SM an internal global symmetry, {\it e.g.} $U(1)_{B-L}$, where $B$ stands for baryon number, and $L$ for lepton number.\footnote{$U(1)_{B-L}$ is not anomalous, meaning it is not broken by non-perturbative quantum mechanical effects, unlike $U(1)_B$ or $U(1)_L$.} Note that, in the massless-neutrino SM, $U(1)_{B-L}$ is an {\sl accidental} global symmetry, {\it i.e.}, it arises as a consequence of the imposed gauge symmetries and the particle content. Among other things, this means that there was, {\it a priori}, no reason to believe that it needed to be conserved by allowed SM extensions, including SUSY, quantum gravitational effects, etc. If Eq.~(\ref{eq:dirac}) is indeed the correct description of neutrino masses, $U(1)_{B-L}$ needs to be ``upgraded'' to an imposed fundamental global symmetry, and it is expected to be preserved by, say, quantum gravity, etc.\footnote{There are ways to officially break lepton number at some very high energy scale and ending up with ``naturally Dirac'' neutrinos.\cite{tony} It may, however, prove impossible to verify that $U(1)_{B-L}$ is violated via any moderately realistic physical process.}

Several distinct mechanisms for explaining naturally light Dirac neutrino masses exist in the various models with either ``large'' or ``warped'' extra dimensions.\cite{add_nus,rs_nus,fat_nus} This is rather convenient, given that most realizations of these models have a rather low ultraviolet cutoff, meaning that these models are effective field theories that need to be replaced by (unknown, and often ``gravity related'') new physics at energy scales well below the Planck mass (more often, close to the electroweak breaking scale). Therefore, in order to avoid generic dimension-five operators as in Eq.~(\ref{eq:majorana}) suppressed by $M\gtrsim 1$~TeV and hence unacceptably large Majorana neutrino masses, one is required to impose $U(1)_{B-L}$ (or something similar) as a fundamental global symmetry.\footnote{In all fairness, this is not a unique ``feature'' of extra-dimensional theories. The MSSM, for example, runs into similar problems if R-parity is not imposed as a fundamental symmetry (or unless one can explain why R-parity violating effects are suppressed, as alluded to in the end of the last section). This is, of course, a consequence of the fact that $U(1)_{B-L}$ is an accidental symmetry of the SM, as discussed above.} I'll briefly present a few different ideas that lead to naturally light Dirac neutrinos.

In scenarios with large, flat extra dimensions, the smallness of the neutrino mass can be explained by postulating that while the SM gauge and matter fields are confined to 3+1 space-time dimensions, the right-handed neutrinos freely propagate in $n+1$ space-time dimensions, similar to the graviton. If this is the case, the four-dimensional effective $LHN$ couplings are ``volume suppressed:
\begin{equation}
y_{i\alpha}\sim y_{i\alpha}^{\delta-\rm dim}\left(\frac{M_{d}}{M_{\rm Pl}}\right)^{\frac{\delta}{d}},
\end{equation}
where $3+\delta$ is the number of extra space dimensions accessible to the $N_i$, $3+d$ is the total number of ``large'' extra dimensions and $M_d$ is the higher dimensional Planck constant. For details, see, for example, Refs.~\refcite{add_nus,add_clfv}. Besides ``predicting'' that the neutrinos are Dirac fermions, these types of models provide several other interesting phenomenological consequences. For example, in the case $\delta=1$ (disfavored for several reasons) the Kaluza-Klein excitations of the $N_i$ mix significantly with the ordinary neutrinos, behaving as a large number of sterile states. Perhaps more interesting are the combined effects of all Kaluza-Klein modes (present for all $\delta$) to charged lepton flavor violation, short-baseline neutrino flavor change, the anomalous magnetic moment of the muon, etc.\cite{add_clfv}

Another scenario that takes advantage of the extra-dimensions in order to explain the physics of flavor is to postulate the existence of a ``fat-brane,'' a $(3+d)+1$-space-time with $d$ relatively small (inverse TeV-size or smaller) compact extra-dimensions where the SM bosons are free to propagate while the SM fermions are confined, somehow, to {\sl distinct} three-dimensional subspaces.\cite{fat_nus} For concreteness, assume one extra-dimension and that all fermion wave-functions in the fifth dimension are well approximated by Gaussians centered at distinct locations $y_f$. This being the case, it is easy to show that the effective four dimensional Yukawa coupling, say, $LHe$, is proportional to $\hat{y}e^{\mu^2(y_L-y_e)^2}$, where $\hat{y}$ is the (dimensionless) five-dimensional Yukawa coupling and $\mu$ is the typical inverse width of the Gaussian profile. The Gaussian dependency of the effective couplings allows one to explain all quark, charged lepton,\cite{mirabelli_schmaltz} and neutrino masses\cite{our_fat} with order one values of $\hat{y}$ and $y_f$ (measured in units of $1/\mu$). There are several interesting phenomenological consequences of fat-brane scenarios, as discussed, for example, in Ref.~\refcite{fat_pheno}.

\section{Lepton Mixing versus Quark Mixing}
\label{section:mixing}

As already alluded to in section \ref{intro}, another theoretical question that has received a significant amount of attention is the fact that the lepton mixing matrix is qualitatively different from the quark one. 

The quark mixing matrix, or the CKM matrix $V_{\rm CKM}$ is, at zeroth order, equal to the identity matrix -- quark mass (or strong) eigenstates are almost identical to weak (or flavor) ones. Furthermore, the off-diagonal entries of $V_{\rm CKM}$ are hierarchical. Schematically,
\begin{equation}
\left|V_{\rm CKM}\right|\sim\left(\begin{array}{ccc} 1 & \epsilon & \epsilon^3 \\ 
\epsilon & 1 & \epsilon^2 \\ 
\epsilon^3 & \epsilon^2 & 1 \end{array}\right), ~~~ \epsilon\sim0.2.
\label{CKM}
\end{equation}
For more quantitative details, see, for example, Ref.~\refcite{pdg}.

The lepton mixing matrix, sometimes refereed to as the MNS matrix $V_{\rm MNS}$,\footnote{$V_{\rm MNS}$ was referred to as $U$ in section~\ref{intro}.} as discussed in section \ref{intro}, is characterized  by two large mixing angles, and most of its entries are of the same order of magnitude. Schematically,
\begin{equation}
\left|V_{\rm MNS}\right|\sim\left(\begin{array}{ccc} 1 & 1 & \varepsilon \\ 
1 & 1 & 1 \\ 
1 & 1 & 1 \end{array}\right), ~~~ \varepsilon\lesssim 0.2.
\label{MNS}
\end{equation}
It is quite apparent that that Eq.~(\ref{MNS}) does not resemble Eq.~(\ref{CKM}) at all, with the possible exception that $\epsilon$ is approximately the same as the upper bound for $\varepsilon$. 

The suspiciously ordered form of Eq.~(\ref{CKM}), together with the fact that the quark masses are very hierarchical ({\it cf.} Fig.~\ref{allmasses}), has received a huge amount of attention over the past thirty years or so. We interpret quark masses and mixing as evidence of some new organizing principle (symmetry) of Nature, which is perhaps broken at inaccessibly high energy scales. This means that the small mixing angles in the CKM matrix may be indicative of the existence of, say, new ``flavor'' quantum numbers capable of distinguishing a top-quark from a charm-quark from an up-quark. 

The question one would like to address is whether the same should apply in the lepton sector, {\it i.e.}, is there evidence that Nature distinguishes, at a more fundamental level, between electrons and muons and taus? There are several indirect reasons to naively believe that this should be the case. First of all, charged lepton masses are hierarchical. Second of all, there are several indications that quarks and leptons may turn out to be different ``low-energy'' manifestations of the same fundamental matter field, as is predicted by most grand unified theories (GUTs). 

GUT relations have lead to the first prediction for the lepton mixing matrix: $V_{\rm MNS}\simeq V_{\rm CKM}$, which has clearly been falsified by the neutrino oscillation data. More recently, several GUT neutrino mass models have succeeded in properly accommodating all neutrino oscillation data.\cite{mixing_revs} The challenge has now shifted from trying to understand whether GUTs can explain large neutrino mixing to sorting out the different possibilities.

Even in the absence of any guidance from the quark sector, symmetry-explanations to neutrino masses and lepton mixing have been heavily undertaken. These are driven by peculiar features of neutrino mixing, such as 
\begin{itemize}
\item some (or all) neutrino masses could be quasi-degenerate in absolute value;
\item $|U_{e3}|$ could be much smaller than all other entries in the MNS matrix;
\item $|U_{\mu3}|$ could be very close to $|U_{\tau3}|$ (maximal atmospheric mixing),
\end{itemize}
and are summarized in the literature.\cite{mixing_revs} Perhaps the most relevant feature of all lepton flavor models (with or without grand unification) is that they can be falsified by precision neutrino oscillation experiments. Indeed, the values of the currently unknown neutrino oscillation observables, spelled out in detail in section \ref{intro} ($|U_{e3}|$, the mass-hierarchy, etc), are predicted by the different lepton flavor models (see, for example, table~1 in Ref.~\refcite{reactor_white}).

It may, however, turn out that we have been asking the wrong question when it comes to neutrino mixing. Upon closer inspection, while Eq.~(\ref{CKM}) looks very structured, Eq.~(\ref{MNS}) looks quite structureless and ``ordinary.'' It has been pointed out,\cite{anarchy} and studied in detailed,\cite{haba_murayama,degouvea_murayama} that Eq.~(\ref{MNS}) is a good representative of a random three-by-three unitary matrix distribution, indicating, perhaps, that, unlike the quark sector, mixing in the lepton sector is ``anarchical'' -- all `peculiar features' mentioned above could turn out to be accidental!

The anarchical hypothesis, spelled out in detail in Ref.~\refcite{degouvea_murayama}, is, contrary to naive expectations, quite consistent with hierarchical charged leptons, very hierarchical neutrino masses,\cite{haba_murayama,degouvea_valle} and grand-unified theories.\cite{murayama_masiero} More interestingly, the anarchical hypothesis is falsifiable -- anarchy prefers values of $|U_{e3}|^2$ not too far from the current upper bound. As a matter of fact, at the $95\%$ confidence level, anarchy predicts $|U_{e3}|^2>0.01$, which will be probed definitively by next-generation neutrino experiments (see, for example, Refs.~\refcite{reactor_white,t2k_nova}). Hence, if $|U_{e3}|^2\ll0.01$, we will be lead to strongly believe that there is indeed structure in lepton mixing. 

The biggest potential criticism against the anarchical hypothesis today is the fact that the atmospheric mixing seems to be maximal: $\sin^22\theta_{23}=1$ is preferred by the current atmospheric and accelerator neutrino data. Maximal mixing implies that the $\nu_3$ mass eigenstate is ``composed'' of equal amounts of $\nu_{\mu}$ and $\nu_{\tau}$: $|U_{\mu3}|^2=|U_{\tau3}|^2$. It would be rather surprising if this turned out to be the case without a ``fundamental reason'' (symmetry), in spite of the fact that a statistical test of anarchy ``prefers'' maximal atmospheric mixing.\cite{degouvea_murayama}

Currently, in spite of the fact that $\sin^22\theta_{23}=1$ is preferred, the atmospheric mixing angle need not be particularly close to maximal at all. The uncertainty on $\cos2\theta_{23}$, a good measure of maximal atmospheric mixing,\cite{theta_23_mine} is very large -- $\cos2\theta_{23}\in[-0.36,0.36]$ at the three sigma confidence level. The key question that can be addressed in practice is how close to zero should $\cos2\theta_{23}$ be if  $\cos2\theta_{23}=0$ is required by some approximate symmetry.\footnote{It is quite unlikely that any flavor symmetry related to maximal atmospheric mixing is exact. Indeed, flavor symmetries usually need to be broken in order to properly fit the charged lepton and the neutrino observables.\cite{mixing_revs}} An attempt at studying this issue can be found in \refcite{theta_23_mine}. There, I analyze different neutrino mass textures that yield, at zeroth order, maximal atmospheric mixing, and discuss the consequences for $\cos2\theta_{23}$ if the zeroth order results were perturbed in a generic way.\cite{theta_23_exceptions} The following results are obtained: (i) $|\cos2\theta_{23}|\sim|U_{e3}|$, {\it i.e.}, the deviation of $\theta_{23}$ from maximal is correlated with the size of $|U_{e3}|$, (ii) $\cos2\theta_{23}\sim \sqrt{\frac{\Delta m^2_{12}}{\Delta m^2_{13}}}\sim 0.1$ in the case of a normal mass-hierarchy, while $\cos2\theta_{23}\sim\frac{\Delta m^2_{12}}{\Delta m^2_{13}}\sim 0.01$, {\it i.e.}, depending on the mass hierarchy, one expects either a reasonably large deviation of $\theta_{23}$ from maximal, to which next-generation experiments are sensitive,\cite{test_maximal} or a tiny deviation from maximal mixing, which will prove very challenging to observe experimentally. 

\section{Conclusions}
\label{conclusions}

New physics beyond the standard model of electroweak and strong interactions has manifested itself in the form neutrino masses. Their effects were observed in several distinct neutrino oscillation experiments.

While we are able to phenomenologically describe neutrino masses and neutrino mixing, there is too little available information to decide what is the fundamental physics behind neutrino masses. In particular, we do not yet know what the correct ``new standard model'' Lagrangian is supposed to be, as discussed in sections 2 and 3. The reason for this is that, for example, Eqs.~(\ref{eq:majorana}) and (\ref{eq:dirac}), while qualitatively distinct, lead to the same observable phenomena, with one very important exception -- the faith of lepton number conservation.\footnote{For the experts, baryon number minus lepton number.} The determination of the whether lepton-number is conserved will help guide our understanding of the physics behind neutrino masses in a most fundamental way.

The most precise probe of lepton number conservation is the search for neutrino-less double beta decay.\cite{0nubb} In a nutshell, one searches for lepton-number violating nuclear decays of the type
\begin{equation}
^AZ\to^A[Z+2]e^-e^-,
\end{equation} 
where $^AZ$ is a nucleus with atomic number $Z$ and mass number $A$. In the presence of massive Majorana neutrinos, the process above happens with a finite lifetime. Furthermore, the amplitude $A_{0\nu\beta\beta}$ for this process is directly proportional to an effective neutrino mass $m_{ee}$, given by
\begin{equation}
m_{ee}=\left|\sum_i U_{ei}^2m_i\right|.
\end{equation} 
$m_{ee}$ depends on the relative complex phases of $U_{ei}$, and is, in principle, sensitive to (a linear combination of) Majorana and Dirac phases. This also means that the different terms in the sum can ``add destructively,'' leading to a suppressed rate for neutrino-less double beta decay. Current experiments\cite{pdg} rule out $m_{ee}\gtrsim 1$~eV, while a controversial and yet-to-be-confirmed analysis of the Heidelberg-Moscow data\cite{Klapdor} claims to have observed lepton number violation associated with $m_{ee}\in[0.24~\rm eV,0.58~eV]$ at the three sigma level.  

Expectations from neutrino oscillation experiments depend on the neutrino mass hierarchy and on precise knowledge of the neutrino oscillation parameters. If neutrinos are Majorana particles and $\Delta m^2_{13}<0$, $m_{ee}\gtrsim 0.05$~eV, while if $\Delta m^2_{13}>0$, $m_{ee}$ could be much smaller than 0.005~eV. While very challenging, it is expected that we will be able to determine whether the neutrinos are Majorana or Dirac fermions with a little luck, a lot of patience, and a significant amount of investment on neutrino-less double beta decay experiments. Several different proposals are currently under investigation, and next-generation experiments are expected to be ready to provide results in a handful of years.\cite{future_0nubb} 

Several ``non-neutrino'' experiments also potentially offer information capable of sorting out the physics behind neutrino masses. Among these, searches for rare charged-lepton processes have received a lot of attention. As alluded to in previous sections, searches for charged-lepton flavor violating processes may prove invaluable for understanding the origin of neutrino masses. In the near/intermediate future, experiments sensitive to $\mu\to e\gamma$\cite{meg} and $\mu\to e$-conversion in nuclei\cite{meco} are expected to start collecting data, while new facilities may provide optimal capability for pursuing these types of endeavors.\cite{future_meg_et_al}

Another source of information is the fact that the lepton mixing matrix is qualitatively different from the quark mixing matrix. This poses a challenge to grand unified theories and may ultimately prove to be one of only a few observables capable of distinguishing different models for grand unification. It may also be interpreted as evidence that the symmetry structure of leptons is distinct from that of quarks, or, ultimately, it may prove to be evidence that there is no symmetry that distinguishes, at a fundamental level, the different neutrino species. In order to start sorting out the different possibilities, it is crucial to determine whether $|U_{e3}|^2$ is larger than 1\%, and whether $\cos2\theta_{23}$ is larger than 10\%.

Finally, I would like to emphasize that the discovery of the neutrino masses and leptonic mixing may turn out to be just the ``tip of the iceberg.'' There are several reasons to suspect (or at least hope) that neutrino oscillation experiments will reveal more about Nature than they already have. 

One is the LSND anomaly. The LSND experiment measured the neutrino flux produced by pion decay in flight ($\pi^+\to\mu^+\nu_{\mu}$) and antimuon decay at rest ($\mu^+\to e^+\nu_e\bar{\nu}_{\mu}$), and observed a small electron-type antineutrino flux some 30 meters away from the production region.\cite{lsnd} The originally absent $\bar{\nu}_e$-flux can be interpret as evidence that $\bar{\nu}_{\mu}$ is transforming into $\bar{\nu}_{e}$ with $P_{\bar{\mu}\bar{e}}$ of the order a fraction of a percent. If interpreted in terms of neutrino oscillations, the LSND anomaly points to a mass-squared difference $\Delta m^2_{\rm LSND}\sim 1$~eV$^2$ (more importantly,  $\Delta m^2_{\rm LSND}\gg10^{-3}$~eV$^2$). The LSND result will be confirmed or refuted, hopefully unambiguously, by the on-going MiniBooNE experiment.\cite{miniboone}

It is easy to understand why the LSND anomaly does not ``fit'' in the three flavor mixing scheme described in section \ref{intro}. With three neutrinos, one can define only two independent mass-squared differences, and these are already ``taken'' by the solar, atmospheric, reactor and accelerator data, and are both much smaller than $\Delta m^2_{\rm LSND}$. Given that LSND results have not yet been confirmed by another experiment -- indeed, the Karmen 2 experiment,\cite{karmen} which could have confirmed the LSND anomaly, did not observe a signal, ruling out a significant portion of the LSND allowed parameter space -- it is more widely believe that  ``ordinary'' three-flavor oscillations are responsible for all-but-LSND-data, while the LSND anomaly could be due to more exotic new physics.

One possible solution to the LSND anomaly is to add extra, SM singlet, neutrinos, capable of mixing with the ordinary, or active, neutrinos. While this allows one to define at least three mass-squared differences, it is not guaranteed that one is capable of fitting all neutrino data with four (or more) neutrino mixing. Indeed, detailed analyses\cite{valle_new,strumia_sterile} suggest that four neutrino mixing schemes are either very poor or at best mediocre fits to all neutrino data. Five neutrino mixing schemes,\cite{five} however, seem to fit the data properly, but some worry that the choices for mixing parameters are rather finely tuned.

More exotic solutions to the LSND anomaly have been proposed, and none of them seem to fit all data particularly well. For example, the possibility that there are rare $\mu\to e\nu_e\bar{\nu}_e$ decays\cite{babu_pakvasa} is disfavored by detailed analysis of the Karmen data,\cite{karmen_mu} while postulating that neutrinos and antineutrinos have different masses and mixing angles\cite{cptv} has been strongly disfavored by KamLAND and atmospheric data.\cite{no_cptv} It is fair to say that if MiniBooNE confirms the observations made by LSND, a detailed analysis of all available data plus potentially new data will be required in order to properly understand the mechanism behind the LSND neutrino flavor change.

Another hint that neutrino experiments may reveal new phenomena besides masses and mixing is the fact that the neutrino oscillation phenomenon is very sensitive to certain very small effects.\cite{nu2004} Unprecedented sensitivity comes from the Òquantum interferometricÓ nature of the oscillation phenomena. It is always useful to keep in mind that neutrino oscillations have allowed us to observe the neutrino masses themselves, and that these may be a manifestation of physics at otherwise inaccessible energy scales.

We have only just begun to explore the physics behind neutrino masses. Much more information is expected in the following several years. While it is unclear at the moment what will be revealed, it is certain that neutrinos have not yet exhausted their ability to surprise the physics community and shape our understanding of Nature.

\section*{Acknowledgments}

This brief review is based on parts of seminars presented at various institutions, and on parts of an invited plenary talk presented at the 2004 meeting of the Division of Particles and Fields, in the University of California at Riverside. I thank all my hosts for their hospitality and too many people to list here for useful conversations. I also thank K.K.~Phua for inviting me to submit this contribution to Modern Physics Letters {\bf A}.

\end{document}